\documentclass[12pt]{article}
\usepackage{tikz}
\usetikzlibrary{calc}
\usepackage{calc,graphicx, complexity}
\usepackage{bez123,calc,curves,ebezier,epic,eepic,graphicx,multiply,rotating}
\everymath{\displaystyle}
\usepackage{graphicx}
\usepackage{pst-all}
\usepackage{color}
\usepackage{amssymb}
\usepackage{amsmath}
\newtheorem{prethm}{{\bf Theorem}}

\newenvironment{thm}{\begin{prethm}{\hspace{-0.5
				em}{\bf .}}}{\end{prethm}}

\newtheorem{prelemma}{{\bf Lemma}}

\newenvironment{lemma}{\begin{prelemma}{\hspace{-0.5
				em}{\bf .}}}{\end{prelemma}}

\newtheorem{preex}{{\bf Example}}

\newtheorem{preprop}{{\bf Proposition}}

\newenvironment{prop}{\begin{preprop}{\hspace{-0.5em}{\bf .}}}{\end{preprop}}

\newtheorem{precor}{{\bf Corollary}}

\newtheorem{preobservation}{{\bf Observation}}

\newenvironment{observation}{\begin{preobservation}{\hspace{-0.5
				em}{\bf .}}}{\end{preobservation}}
\newtheorem{preremark}{{\bf Remark}}

\newtheorem{preprob}{{\bf Problem}}

\newtheorem{predefin}{{\bf Definition}}

\newenvironment{defin}{\begin{predefin}{\hspace{-0.5
				em}{\bf .}}}{\end{predefin}}

\newtheorem{preconj}{{\bf Conjecture}}

\newtheorem{preprobb}{{\bf Problem}}

\newtheorem{prelem}{{\bf Theorem}}

\newenvironment{proof}{{\bf Proof.}\rm }{\hfill{$\Box$}}

\newtheorem{presolution}{{\bf Solution.}}

\def\newpic#1{}

\title{\vspace{-2.5cm}\Large\bf More algorithmic results for problems of spread of influence in edge-weighted graphs with and without incentives}

\author{\large\bf Siavash Askari~~~~~Manouchehr Zaker\footnote{mzaker@iasbs.ac.ir}
	\vspace{5mm}\\
	Department of Mathematics,\\
	Institute for Advanced Studies in Basic Sciences,\\
	Zanjan 45137-66731, Iran} \date{}
\begin{document}	
\maketitle
\vspace*{-1cm}
\begin{abstract}
\noindent Many phenomena in real world social networks are interpreted as spread of influence between activated and non-activated network elements. These phenomena are formulated by combinatorial graphs, where vertices represent the elements and edges represent social ties between elements. A main problem is to study important subsets of elements (target sets or dynamic monopolies) such that their activation spreads to the entire network. In edge-weighted networks the influence between two adjacent vertices depends on the weight of their edge. In models with incentives, the main problem is to minimize total amount of incentives (called optimal target vectors) which can be offered to vertices such that some vertices are activated and their activation spreads to the whole network. Algorithmic study of target sets and vectors is a hot research field. We prove an inapproximability result for optimal target sets in edge weighted networks even for complete graphs. Some other hardness and polynomial time results are presented for optimal target vectors and degenerate threshold assignments in edge-weighted networks.
\end{abstract}


	
\section{Introduction and related models}

\noindent Spread of influence is a process in which individuals in a virtual or real world community change their opinions or any kind of influence through communication and interaction with each other. Various phenomena are spreading in real and virtual social networks, in which the members of the network who are connected are affected. Adaptation of new economical products by word-of-mouth communication is one example. Also in elections people usually decide to vote or not to vote based on the influence of other people in their neighborhood. Viral marketing is another phenomenon, which refers to the dissemination of information in products, behaviors, and its acceptance by people in the community. Therefore, the ability to control the spread of these phenomena is economically and politically desirable, which has created many optimization problems. Let $G=(V, E)$ be a graph representing a social network in which vertices represent the network members and edges indicate the social tie or relationship between the members of the network. We denote the set of neighbors and the vertex degree of every vertex $v$ by $N(v)$ and $d_G(v)=\left| N(v)\right|$, respectively. Throughout the paper, by a threshold assignment for an underlying network $G$, we mean any function $\tau: V\longrightarrow \mathbb{N}$ which assigns thresholds to the vertices such that $1\leq \tau(v)\leq d_G(v)$, for all $v\in V$. The value $\tau(v)$ indicates the hardness of susceptibility or influenceability of vertex $v$ in front of an influence.

\noindent The activation process corresponding to $(G,\tau)$ is defined as follows. Firstly, a subset $A_0$ of vertices in $G$ is activated. Denote by $A_t$ the set of active vertices in each round $t$. Then a vertex $x\in V(G)\setminus A_t$ becomes active in round $t+1$ if and only if vertex $x$ has at least $\tau(x)$ neighbors in $A_t$. Note that once a vertex is activated it remains active until the end of process. Such a subset $A_0$ is called target set of dynamic monopoly in $(G,\tau)$ if it activates the entire graph. The smallest cardinality of dynamic monopolies in $(G,\tau)$ is denoted by $dyn(G,\tau)$. The target set selection problem (TSS) is a decision problem which for any instance $(G, \tau)$ and integer $k$, asks wether $dyn(G,\tau)\leq k$. Target set selection and dynamic monopolies have been investigated by various authors  \cite{ABW,CL,Ch,DR,FKRRS,KSZ1,Z1}. Many algorithmic and hardness results for TSS were obtained in
\cite{Ch,CNNW,CGMRV,FK}. By TSS$(d=3,t\in \{1,2\})$ they mean the TSS problem restricted to regular graphs of degree $3$, where all thresholds are $1$ or $2$. The following is Theorem 1.8 in \cite{FK}.

\begin{thm}(\cite{FK})
The TSS$(d=3,t\in \{1,2\})$ problem cannot be approximated within the ratio
of $\mathcal{O}(2^{\log^{1-\epsilon}n})$ for any fixed constant $\epsilon >0$, unless $\P=\NP$.\label{feige}
\end{thm}

\subsection{Target set selection in edge-weighted networks}

\noindent Two different weighted versions of TSS were investigated in the literature. The models with weighted vertices
and with weighted edges were studied in \cite{RZ} and \cite{CCGMPV}, respectively. If an edge $e=uv$ has weight $w(e)$
it means that the amount of influence between $u$ and $v$ is $w(e)$. The networks with weighted edges are more
realistic than non-weighted networks. In case that the underlying network is edge-weighted, by $\omega$ we always mean
a weight function which assigns a non-negative rational number to each edge of $G$. No need to assume irrational weights since they can be approximated by rational numbers. But sometime we insist that the weights are positive rational numbers. The spread of influence in this case naturally depends on the weight of the edges. The formal definition, presented in \cite{Z2} is given in the following.

\begin{defin}
Let a triple $(G,\omega,\tau)$ be given, where $G$ is a simple graph and $\tau$ and $\omega$ are threshold and weight functions for the vertices of $G$, respectively. The activation process corresponding to $(G,\omega,\tau)$ is defined as follows. Initially a subset $A_0$ of vertices in $G$ is activated. $A_t$  is the set of vertices activated in round $t$. Then a vertex $x\in V(G)\setminus A_t$ becomes active in round $t+1$ if and only if the following inequality holds,
where $E_{t}(x)$ consists of all edges say $e$ such that $e=xy$ for some vertex $y\in A_{t}$.
\begin{align*}
\sum\limits_{e\in E_{t}(x)}\omega(e)\geq\tau(x).
\end{align*}
Such a set $A_0$ is called dynamic monopoly in $(G,\omega,\tau)$ if by activating the vertices in $A_0$, the entire graph get activated. Denote the smallest cardinality of dynamic monopolies by $dyn(G,\omega,\tau)$ and call it an optimal target set of $G$.
\end{defin}

\noindent The target set selection problem for edge-weighted graphs is defined as follows.

\noindent \textbf{Name:} Target Set Selection in (edge)-Weighted graphs (TSSW).\\
\noindent \textbf{Instance:} A triple $(G, \tau, \omega)$.\\
\noindent \textbf{Goal:} Find $dyn(G,\omega,\tau)$.

\subsection{Target vectors and spread of influence with incentives}

\noindent To define the model with incentives, we start with a simple and practical example. Consider a company that wants to sell its products. The company may decide to use a discount on its products instead of offering a few products for free (as an award) in order to encourage people to buy these products. In fact, if members of the community are considered as vertices of the network, in this method, instead of focusing on $A_0\subseteq V(G)$ as the target set, we target all vertices of the network so that the entire network is encouraged (or activated). In other words, we consider discounts as incentives on each vertex, and the goal is to activate the entire network by minimizing the sum of incentives assigned to the vertices. These ideas are introduced in \cite{DHMMRSZ}. Then Cordasco et al. \cite{CGRV} formalized the related model as follows.

\noindent Let $G=(V, E)$, where $V=\{v_1, v_2, \ldots, v_n\}$ represent an underlying network and $\tau$ be an assignment of thresholds to the vertices of $G$. An assignment of incentives to the vertices of $V$ is a vector $\mathbf{p}=\left(p(v_{1}), \cdots, p(v_{n}) \right)$, where $p(v)\in \{0,1,\ldots\}$ represents the amount of incentive we apply/consume on $v\in V$.
%The arbitrary vertex $v$ is activated when $p(v)+ \left| active[N(v)]\right| \geq \tau(v)$.
It can be assumed that $0\leq p(v)\leq\tau(v)\leq d(v)$. The activation process in $G=(V, E)$, starting from the incentives $\mathbf{p}$ is as follows
\begin{align*}
active[\mathbf{p},0]=\left\lbrace v \mid \ p(v)\geq \tau(v)\right\rbrace.
\end{align*}
Then, for all $t\geq 1$ define:
\begin{align*}
active[\mathbf{p},t]=active[\mathbf{p},t-1]\cup \left\lbrace u:\left|N(v)\cap active[\mathbf{p},t-1] \right|\geq \tau(u)-p(v) \right\rbrace.
\end{align*}

\noindent A vertex $v$ is activated at round $t>0$ if $v\in active[\mathbf{p},t]\setminus active[\mathbf{p},t-1].$
An assignment of incentives $\mathbf{p}$ is called a target vector whenever the activation process activates the entire network vertices with this assignment, that is, $active[\mathbf{p},t]=V$ for some $t\geq 0.$ The size of the incentive assignment $\mathbf{p}:V\longrightarrow \mathbb{N}\cup \{0\}$ is given by ${\sum}_{v\in V} p(v)$. A target vector of the minimum size is called an optimal target vector and is denoted by $\mathbf{p}^{\ast}$. The target set selection with incentives or optimal target vector problem is as follows.

\noindent\textbf{Name:} Optimal Target Vector (OTV).\\
\textbf{Instance:} A graph $G=(V,E)$ with a threshold assignment $\tau:V\longrightarrow \mathbb{N}$.\\
\textbf{Goal:} Find a target vector $\mathbf{p}$ with minimum possible value $\mathbf{p}(V)=\sum\limits_{v\in V}p(v).$

\noindent Comprehensive studies have focused on this problem with different titles and obtained many results \cite{CGPRV,CGRV,ER,GRZ,MVMO}. Using the result obtained by Chen \cite{Ch}, Cordasco et al. proved that OTV cannot be approximated within a ratio of $O(2^{\log ^{1-\varepsilon}n})$, for any fixed $\varepsilon>0$, unless $\NP\subseteq \DTIME\left(n^{polylog(n)} \right)$ \cite{CGRV}. Feige and Kogan proved that OTV has polynomial time solution if the threshold of any vertex in the input graph is $1$ or $d(v)$, and or $\tau(v)\in\left\lbrace d(v)-1, d(v)\right\rbrace$ \cite{FK}. OTV is also solvable in polynomial time for complete graphs, trees, and cycles \cite{CGRV}. OTV has also been investigated by G{\"u}nne{\c{c}} et al. for bipartite graphs and directed graphs and proved to be solvable for directed
acyclic graphs in polynomial time \cite{GRZ,GRZ2}.

\noindent We know that networks with weighted edges are more realistic in the real world networks. For example, in the advertisement for adaptation of new products, very influential persons have more activation effects than the ordinary people. Therefore, we quantify these influences as weight of the edges between participants in the community so that the spread of influence depends on the weight of the edges. In the setting of G{\"u}nne{\c{c}} et al. \cite{GRZ} (under the name of least-cost influence maximization problem) influence between any two vertices $u$ and $v$ depends on weight of
the edge $e=uv$. In the following, using the model of target set selection with incentives presented in \cite{CGRV} and the TSSW model given by the second author \cite{Z2}, we present our model for spread of influence with incentives in edge-weighted graphs.

\noindent Consider a quadratic $(G,\omega,\tau,\mathbf{p})$ in which $G$ is a simple graph with a weight function $\omega:E(G)\rightarrow \left[ 0,\infty\right)$ and a threshold assignment $\tau:V\rightarrow \left[ 0,\infty\right)$ and an incentive assignment $\mathbf{p}:V\rightarrow \left[ 0,\infty\right)$. The activation process corresponding to $(G,\omega,\tau,\mathbf{p})$ is defined as follows. Denote by $p(v)$ the incentive received by every vertex $v$. Initially a subset $A_0$ of vertices in $G$ is activated. Precisely, $A_0=\{v: p(v)\geq\tau(v)\}$. Define $A_t$ as a set of vertices activated in round $t$. Then a vertex $x\in V(G)\setminus A_t$ becomes
active in round $t+1$ if and only if the following inequality holds, where $E_t(x)$ consists of all edges say $e$ such that $e=xy$ for some vertex $y\in A_t$.

\begin{align*}
\sum\limits_{e\in E_{t}(x)}\omega(e)+p(x)\geq\tau(x).
\end{align*}

\begin{defin}
Given $(G,\omega,\tau)$, by a target vector $\mathbf{p}$ we mean an incentive assignment $\mathbf{p}$ such that the activation process corresponding to $(G,\omega,\tau,\mathbf{p})$ activates the whole graph $G$.
\end{defin}

\noindent The weighted version of OTV for networks with weighted edges, denoted by OTVW, is the following.

\noindent \textbf{Name:} Optimal Target Vector in Weighted Graphs (OTVW).\\
\noindent \textbf{Instance:} A triple $(G,\tau,\omega)$, $\tau$ a threshold assignment and $\omega$ a weight function.\\
\noindent \textbf{Goal:} Find a target vector $\mathbf{p}$ which minimizes $\mathbf{p}(V)=\sum\limits_{v\in V}p(v)$.

\subsection{Degenerate threshold functions}

\noindent The degenerate threshold functions was defined by Feige and Kogan \cite{FK} as follows. Let $G$ be a simple undirected graph. A threshold function $\tau$ is a degenerate assignment for vertices of $G$ if in every induced subgraph $H$ of $G$, there exists a vertex $x\in V(H)$ such that $\tau(x)\geq d_H(x)$. This notion is similar to the concept
of generalized degeneracy defined in \cite{Z12}. Given $\kappa: V(G)\rightarrow \mathbb{N}\cup \{0\}$, a graph $G$ is
called $\kappa$-degenerate if the vertices of $G$ can be ordered as $v_1, v_2, \ldots, v_n$ such that
$d_{G[v_1, \ldots, v_i]}(v_i)\leq \kappa(v_i)$, for any $i\in \{1, \ldots, n\}$. It was proved in \cite{Z12} that a set $D$ in $(G,\tau)$ is a target set if and only if $V(G)\setminus D$ is $\kappa$-degenerate, where $\kappa(v)=d_G(v)-\tau(v)$.
The TSS problem when the threshold function is degenerate is denoted by TSS(degenrate). The first complexity results concerning degenerate assignments were obtained in \cite{FK}. We generalize the idea for edge-weighted graphs.

\begin{defin}
Given a triple $(G,\omega,\tau)$, where $G$ is a simple graph, a threshold function $\tau$ for the vertices of $G$ is called degenerate if in every induced subgraph $H$ of $G$, there exists a vertex $x\in V(H)$ such that $\tau(x)\geq {\sum}_{e\in E(x,H)}\omega(e)$, where $E(x,H)$ contains all edges of $H$, incident to $x$. The TSSW problem for $(G,\omega,\tau)$, in which $\tau$ is a degenerate threshold function, is denoted by TSSW(degenerate).
\end{defin}

\noindent Feige and Kogan obtained an approximation algorithm for TSS(degenerate) \cite{FK}. By OTVW(degenerate) we mean the problem OTVW such that the threshold functions in its input $(G,\omega,\tau)$ are restricted to degenerate threshold functions. The following observation is derived routinely.

\begin{observation}\label{obse1}
Given $(G,\omega,\tau)$ on $n$ vertices, where $\tau$ is a degenerate threshold assignment, then there is a degeneracy ordering of its vertices $R:u_1, u_2, \cdots, u_n$ such that for each $1\leq i\leq n$, $\tau(u_i)\geq {\sum}_{e\in E(u_i,R)}\omega(e)$, where $E(u_i,R)$ consists of all edges between $u_i$ and $\{u_1, u_2, \cdots, u_{i-1}\}$.
\end{observation}

\subsection{A table of problems and their complexity status}

\noindent The following table summarizes the complexity results concerning the decision problems discussed in this paper.

\setlength{\tabcolsep}{10pt}
\begin{center}
\begin{tabular}{|c|c|c|c|c|c|}
\hline
Problem & Hardness &  References \\
\hline
TSS & inapproximable
%within the ratio of $\mathcal{O}(2^{\log^{1-\epsilon}n})$,
unless $\NP=\P$ & \cite{Ch,DR} \\
\hline
 OTV &  \NP-complete & \cite{CGRV}\\
\hline
OTVW
& \NP-complete for complete graphs &  \cite{AZ} \\
\hline
TSSW & inapproximable for complete graphs unless $\NP=\P$ & This paper \\
\hline
OTV(degenerate) & Polynomial-time & \cite{FK} \\
\hline
TSS(degenerate) & \NP-complete & \cite{FK} \\
\hline
TSSW(degenerate) & \NP-complete for complete graphs & This paper \\
\hline
OTVW(degenerate) & Polynomial-time & This paper \\
\hline
TSSWD & \NP-complete for directed tournaments & This paper \\
\hline
\end{tabular}
\end{center}

\section{Results for TSSW and TSSW(degenerate)}

\noindent In the following by TSSW(complete) we mean the problem TSSW restricted to edge-weighted complete graphs. Theorem \ref{feige} asserts that TSS$(d=3,t\in \{1,2\})$ cannot be approximated within the ratio
of $\mathcal{O}(2^{\log^{1-\epsilon}n})$ for any fixed constant $\epsilon >0$, unless $\P=\NP$. We use this result to prove a same inapproximibility result for TSSW(complete). This in particular shows that TSSW(complete) and then TSSW is $\NP$-hard.

\begin{prop}\label{prop1}
For any $\epsilon>0$, TSSW(complete) does not admit polynomial time approximation algorithm within ratio $\mathcal{O}(2^{\log^{1-\epsilon}n})$, unless $\NP=\P$.
\end{prop}

\noindent \begin{proof}
We obtain a gap-preserving reduction from TSS$(d=3,t\in \{1,2\})$ to TSSW(complete). Let $(G, \tau)$ be an instance of TSS$(d=3,t\in \{1,2\})$ on $n$ vertices. We obtain a complete graph $K_n$ from $G$ such that $V(K_n)=V(G)$ as follows. For each vertex $v\in V(K_n)=V(G)$ define $\tau'(v)=n\tau(v)$. For each edge $e\in E(K_n)$ define $\omega(e)=n$, if $e\in E(G)$ and $\omega(e)=1$, if $e\not\in E(G)$. We prove that any target set $D$ for $(G,\tau)$ is also a target set in $(K_n, \tau', \omega)$ and vice versa.

\noindent Assume that $D_0\subseteq V(G)$ is a target set for $(G, \tau)$ and let $|D_0|=k$. Suppose that the activation process in $G$ has started with $D_0$. Denote by $D_i$ the set of vertices activated up to round $i$ in this activation process. We show that $D_0$ is a target set for $K_n$ as well. Let $v\in V(K_n)\setminus D_0$ be an arbitrary vertex. We prove that if $v$ is activated in $G$ in a round say $i$, then $v$ as a vertex in $K_n$ becomes active in round $i$. Note that $v$ receives $n|E(v,D_{i-1})|+|D_{i-1}|-|E(v,D_{i-1})|$ influence from its neighbors in $K_n$ at round $i$, where $E(v,D_{i-1})$ consists of all edges $e=(v,u)$ in $G$ such that $u\in D_{i-1}$. Hence, it is enough to prove the following inequality.
\begin{equation}\label{eq1.6}
n|E(v,D_{i-1})|+|D_{i-1}|-|E(v,D_{i-1})|\geq \tau'(v)=n\tau(v).
\end{equation}
Vertex $v$ is activated in $G$ at round $i$. Hence, $|E(v,D_{i-1})| \geq \tau(v)$ and
$$n|E(v,D_{i-1})| \geq n\tau(v)=\tau'(v).$$
\noindent Since $|D_{i-1}|-|E(v,D_{i-1})|\geq 0$, therefore inequality \eqref{eq1.6} holds.

\noindent To prove the converse, let $W_0\subseteq V(K_n)$ be a target set for $(K_n,\tau',\omega)$. Set $|W_0|=k$. Suppose the activation process in $K_n$ is started with $W_0$. The set of vertices activated up to round $i$ is denoted by $W_i$, and $t$ is the total number of activation rounds in this graph. We claim that $W_0$ is a target set for $G$. Let the activation process in $G$ be started with $W_0$. We set $D_0=W_0$ and denote the set of vertices activated up to round $i$ in $G$ by $D_i$. By induction on $0<j\leq t$, we show that $D_j=W_j$. In other words, we show that if the arbitrary vertex $v$ of $K_n$ is activated in a round $j$, then it is activated in $G$ in the same round. Suppose that $D_i=W_i$ for each $i<j\leq t$. Then $D_{j-1}=W_{j-1}$. Let $v\in W_j$ be an arbitrary vertex. We show that $v\in D_j$.
Note that $|D_{j-1}|\leq n-1$. Since $v$
is activated in $K_n$ at round $j$ we have the following inequality, where $E(v,D_{j-1})$ consists of all edges $e=(v,u)$ in $G$ such that $u\in D_{j-1}$.
\begin{equation}\label{eq26}
n\tau(v)=\tau'(v) \leq n|E(v,D_{j-1})|+|D_{j-1}|-|E(v,D_{j-1})|.
\end{equation}
Inequality \ref{eq26} together with the clear inequality $|D_{j-1}|-|E(v,D_{j-1})|\leq n-2$ imply
$$n\tau(v) \leq n|E(v,D_{j-1})|+(n-2)$$
$$\tau(v)-|E(v,D_{j-1})| \leq \frac{n-2}{n} <1$$
$$\tau(v)-|E(v,D_{j-1})|  \leq 0$$
$$\tau(v) \leq |E(v,D_{j-1})|.$$
\noindent Hence, $v$ is activated in round $j$ in $G$. Similarly, by induction on $j$, we show that $D_j\subseteq W_j$. Let $v$ be an arbitrary vertex in $D_j$. Hence, $|E(v,D_{j-1})|\geq \tau(v)$ and $n|E(v,D_{j-1})|\geq n\tau(v)$. Also, since $|D_{j-1}|-|E(v,D_{j-1})|\geq 0$, therefore
\begin{equation}\nonumber
n|E(v,D_{j-1})|+|D_{j-1}|-|E(v,D_{j-1})|\geq n\tau(v)=\tau'(v).
\end{equation}
In other words, $v$ in $(K_n,\omega)$ is activated in round $j$. Namely, $v\in W_j$.
\end{proof}

\noindent Let a triple $(G,\omega,\tau)$ be given. Recall from Subsection 3.1 that a threshold function $\tau$ is called degenerate if in every induced subgraph $H$ of $G$, there exists a vertex $x\in V(H)$ such that $\tau(x)\geq {\sum}_{e\in E(x,H)}\omega(e)$, where $E(x,H)$ contains all edges of $H$, incident to $x$. Recall also that TSSW(degenerate) is a subproblem of TSSW, where the threshold functions are degenerate.

\noindent Feige and Kogan obtained an approximation algorithm for TSS(degenerate) \cite{FK}. We obtain an approximation algorithm for TSSW(degenerate) by generalizing their method. Then we show that the problem is $\NP$-complete even for complete graphs. Given $(G,\omega,\tau)$, in the following we mean $\tau_{max}=\max\{\tau(v): v\in V(G)\}$, $OPT(G)$ is the size of an optimal target set in $G$ and
$c={\min}_{u\in R}\{\tau(u)-{\sum}_{e\in E(u,R)} \omega(e): \tau(u)-{\sum}_{e\in E(u,R)} \omega(e) \ne 0\}$, where $R$ is a degeneracy ordering of $V(G)$ corresponding to $\tau$ obtained in Observation \ref{obse1}. Obviously $\tau_{max}/c\geq 1$.

\noindent \textbf{Algorithm $I$}\\
\noindent \textbf{Input}: A triple $(G,\omega,\tau)$ in which $G$ is a simple graph on $n$ vertices, $\omega$ is the weight function on $E$, and $\tau$ is a degenerate threshold assignment for $V$.\\
\textbf{Output}: A target set $S$ of size at most $(\tau_{max}/c)~OPT(G)$.

\noindent $1.$ $S=\varnothing$\\
\noindent $2.$ Let $R:u_1,u_2,\cdots,u_n$ be a degeneracy ordering of the vertices of $G$\\
\noindent $3.$ For each $1\leq i\leq n$, if $\tau(u_i)>{\sum}_{e\in E(u_i,R)}\omega(e)$
then
$S\leftarrow S\cup \{u_i\}$\\
\noindent $4.$ Return $S$

\begin{thm}
Given a triple $(G,\omega,\tau)$ on $n$ vertices, where $\tau$ is a degenerate threshold assignment. Then Algorithm $I$ has running time $\mathcal{O}(n)$ and returns a target set of size at most $(\tau_{max}/c)~OPT(G)$.
\end{thm}

\noindent \begin{proof}
\noindent Let $S$ be an output of Algorithm $I$. Set $S$ is a target set for $G$, because for each $1\leq i\leq n$, if vertex $u_i$ is not selected by the algorithm, then according to the property of degeneracy ordering, we have $\tau(u_i)={\sum}_{e\in E(u_i,R)}\omega(e)$. Therefore, the elements of $S$ activates all vertices not selected by the algorithm, exactly based on the order in which they are in the degeneracy ordering. Let $|S|=s$ we show that $s\leq(\tau_{\max}\times  OPT(G))/c$. Since $R$ is degeneracy order by condition $3$ of the algorithm and by Observation \ref{obse1}, for each vertex $u$, $\tau(u)\geq{\sum}_{e\in E(u,R)} \omega(e)$, then

\begin{align*}
%\sum\limits_{i=1}^{n}\tau(u_i)&>\sum\limits_{u\in S}\left[\sum\limits_{e\in E(u,R)}\omega(e)\right]+\sum\limits_{u\in V(G)\setminus S}\left[\sum\limits_{e\in E(u,R)}\omega(e)\right]\\ \nonumber
\sum\limits_{i=1}^{n}\tau(u_i)& \geq \sum\limits_{u\in S}\left[\sum\limits_{e\in E(u,R)}\omega(e)+(\tau(u)-\sum\limits_{e\in E(u,R)} \omega(e))\right]+\sum\limits_{u\in V(G)\setminus S}\left[\sum\limits_{e\in E(u,R)}\omega(e)\right]
\end{align*}
Recall that
$$c={\min}_{u\in R}\{(\tau(u)-{\sum}_{e\in E(u,R)} \omega(e)): \tau(u)-{\sum}_{e\in E(u,R)} \omega(e) \ne 0\}$$
$$={\min}_{u\in S}\{\tau(u)-{\sum}_{e\in E(u,R)} \omega(e)\},$$
\noindent hence we have
\begin{align*}
\sum\limits_{i=1}^{n}\tau(u_i)&\geq\sum\limits_{u\in S}\left[\sum\limits_{e\in E(u,R)}\omega(e)+c\right]+\sum\limits_{u\in V(G)\setminus S}\left[\sum\limits_{e\in E(u,R)}\omega(e)\right]\\ \nonumber
&\geq sc+\sum\limits_{u\in S}\left[\sum\limits_{e\in E(u,R)}\omega(e)\right]+\sum\limits_{u\in V(G)\setminus S}\left[\sum\limits_{e\in E(u,R)}\omega(e)\right]\\ \nonumber
&= sc+\sum\limits_{u\in V(G)}\left[\sum\limits_{e\in E(u,R)}\omega(e)\right]= sc+\sum\limits_{e\in E(G)}\omega(e).
\end{align*}
Therefore
\begin{equation}\label{equ3}
\sum\limits_{i=1}^{n}\tau(u_i)\geq sc+\sum\limits_{e\in E(G)}\omega(e).
\end{equation}
If $S^{\star}$ is an optimal target set of cardinality $OPT(G)$, then we have
${\sum}_{u\in V(G)\setminus S^{\star}}\tau(u)\leq|E(G)|$
\cite{ABW}. Therefore
\begin{equation}\label{equ4}
\sum\limits_{u\in V(G)\setminus S^{\star}}\tau(u)\leq \sum\limits_{e\in E(G)}\omega(e).
\end{equation}
By inequality \eqref{equ3} we have
\begin{equation}\label{equ3.7}
\sum\limits_{u\in S^{\star}}\tau(u)+\sum\limits_{u\in V(G)\setminus S^{\star}}\tau(u)\geq sc+\sum\limits_{e\in E(G)}\omega(e).
\end{equation}
By inequalities \eqref{equ4} and \eqref{equ3.7}, we conclude $sc\leq \sum_{u\in S^{\star}}\tau(u)$ and
$s\leq \frac{\tau_{max}}{c}~OPT(G)$.
\end{proof}

\noindent In the following we use a result from \cite{Z2}. By a vertex cover in a graph $G$, we mean any subset $B\subseteq V(G)$ such that each edge $e$ has at least one endpoint in $B$. Denote by $\beta(G)$ the minimum $|B|$, where $B$ is a vertex cover in $G$. Given any triple $(G,\omega, \tau)$, it was proved in Proposition 2.4 \cite{Z2} that $(G,\omega, \tau)$ has a dynamic monopoly (target set) of cardinality at most $\beta(G)$. Note that for any connected graph $G$, $\beta(G)\leq |V(G)|-1$. It was proved in Corollary 3.7 of \cite{FK} that TSS(degenerate) is $\NP$-complete. In the following by TSSW(degenerate,complete) we mean the problem TSSW restricted to degenerate threshold assignments, where the underlying graph is complete. We show in the following that TSSW(degenerate,complete) is \NP-complete. We need some information from the proof of Proposition \ref{prop1}. In the proof, corresponding to each instance $(G, \tau)$ of TSS on $n$ vertices, we obtained a complete graph $K_n$ from $G$ such that $V(K_n)=V(G)$, a threshold assignment $\tau'$ such that $\tau'(v)=n\tau(v)$, $v\in V(K_n)$ and a weight assignment $\omega$ such that $\omega(e)=n$ for $e\in E(G)$ and $\omega(e)=1$ for $e\not\in E(G)$. Denote the triple $(K_n, \tau', \omega)$ by $H_n$. Proposition \ref{prop1} proves that every target set $D$ for $(G,\tau)$ is a target set for $H_n$ and vice versa.

\begin{prop}
TSSW(degenerate,complete) is \NP-complete.
\end{prop}

\noindent \begin{proof}
\noindent Clearly TSSW(degenerate,complete) belongs to $\NP$. We obtain a polynomial time reduction from TSS(degenerate) to TSSW(degenerate, complete). Let $(G,\tau)$ be an instance of TSS(degenerate) on vertex set $\{v_1, \ldots, v_n\}$. We make an instance $(K_{n+1}, \omega, \tau'')$ of TSSW(degenerate, complete) as follows.
Add a new vertex $v_{n+1}$ to $G$ and connect it to every vertex in $V(G)$. We have $V(K_{k+1})=V(G)\cup \{v_{n+1}\}$. Define $\tau''$ as follows. For each $i\in \{1, \ldots, n\}$, set $\tau''(v_i)= n\tau(v_i)+n$ also $\tau''(v_{n+1})= n^2$. We define weight assignment for the edges of $K_{n+1}$ as follows. Define $\omega(v_iv_{n+1})=n$, for each $i\in \{1, \ldots, n\}$. Then for any other edge $e\in E(K_{n+1})$, if $e\in E(G)$ then set $\omega(e)=n$, otherwise $\omega(e)=1$.

\noindent We prove that $\tau''$ is degenerate. Since $\tau$ is degenerate in $G$ then by definition there exists an ordering of $V(G)$ such as $R: v_1, v_2, \ldots, v_n$ such that $\tau(v_i)\geq d_R(v_i)$, where $d_R(v_i)$ denotes the
number of neighbors of $v_i$ in $G[v_1, \ldots, v_i]$. It is easily seen that $\tau''$ with ordering of vertices $R': v_1, \ldots, v_n, v_{n+1}$ is degenerate. We only have to check the last vertex $v_{n+1}$.

\noindent We show in the following that $(G,\tau, k)$ is Yes-instance for TSS(degenerate) if and only if $(K_{n+1}, \omega, \tau'', k+1)$ is Yes-instance for TSSW(degenerate, complete). First, let $D$ be a target set for $(G,\tau, k)$. We prove that $D\cup \{v_{k+1}\}$ is a target set in $(K_{n+1}, \omega, \tau'')$. In $K_{n+1}$ after
activating $v_{k+1}$ the threshold of each vertex $v_i$, $1\leq i\leq n$, is reduced by $\omega(v_iv_{k+1})=n$. Hence, for each $v_i$, the new threshold of $v_i$ is $\tau'(v_i)=\tau''(v_i)-n=n\tau(v_i)$. We have also $\omega(e)=n$ for $e\in E(G)$, otherwise $\omega(e)=1$. We observe that $(K_n, \tau', \omega)$ is identical to $H_n$ defined before the proof. From the proof of Proposition \ref{prop1}, we know that $D$ is a target set in $H_n$ since it is target set in $(G,\tau)$. This proves that $D\cup \{v_{k+1}\}$ is a target set in $(K_{n+1}, \omega, \tau'')$ of size $k+1$.

\noindent Assume now that $S$ is a target set of size $k+1$ in $(K_{n+1}, \omega, \tau'')$, where $k\geq 0$. There are two possibilities.

\noindent Case 1. $v_{n+1}\not\in S$.

\noindent In this case, since $\tau(v_{n+1})={\sum}_{i=1}^n \omega(v_iv_{n+1})$, then $v_i\in S$ for each $i$
with $1\leq i \leq n$. Hence, $k+1\geq n$ or $k\geq n-1\geq \beta(G)$. It follows that $G$ has a target set of at most $\beta(G) \leq k$ vertices, as desired.

\noindent Case 2. $v_{n+1}\in S$.

\noindent In this case we prove that $S'=S\setminus v_{n+1}$ is a target set in $G$. Clearly, $S'\subseteq V(G)$ is a target set for $H_n=(K_n, \tau', \omega)$. It follows from the comment before the proof that $G$ admits a target set of size $|S'|=k$. This completes the proof.
\end{proof}

\section{Results for OTVW and OTVW(degenerate)}

\noindent We first present a result for OTVW(degenerate). Feige and Kogan proved that OTV(degenerate) has a polynomial time solution \cite{FK}. We use the following result proved in Proposition 2 in \cite{AZ}.

\begin{prop}(\cite{AZ})\label{propaz}
Let $(G, \omega, \tau)$ be a weighted graph and $\mathbf{p}^{\ast}$ be an optimal target vector in $G$. Then
\begin{align*}
\sum\limits_{v\in V(G)}\mathbf{p}^{\ast}(v)\geq \sum\limits_{v\in V(G)}\tau(v)-\sum\limits_{e\in E(G)}\omega(e).
\end{align*}
\end{prop}

\noindent By generalizing method of \cite{FK}, we show that OTVW(degenerate) is solved in polynomial time.

\begin{prop}\label{prop2}
OTVW(degenrate) can be solved in polynomial time.
\end{prop}

\noindent \begin{proof}
\noindent Let $G=(V,E), \omega, \tau)$ be given, where $\tau$ is degenerate. By Proposition \ref{propaz}
\begin{align}\label{n3.1}
\sum\limits_{v\in V(G)}p^{*}(v)\geq \sum\limits_{v\in V(G)}\tau(v)-\sum\limits_{e\in E(G)}\omega(e).
\end{align}
Since $\tau$ is degenerate, we deduce from Observation \ref{obse1} that there exists an ordering of vertices in $G$ such as $R:u_1,u_2,\cdots,u_n$ such that $\tau(u_i)\geq {\sum}_{e\in E(u_i,R)}\omega(e)$, for each $1\leq i\leq n$. Now, starting from $u_1$ scan the vertices in $R$ and for each $1\leq i\leq n$, if $\tau(u_i)>{\sum}_{e\in E(u_i,R)}\omega(e)$ then by placing $p(u_i)=\tau(u_i)-{\sum}_{e\in E(u_i,R)}\omega(e)$, vertex $u_i$ is activated. Otherwise, $u_i$ is activated by its previous neighbors in $R$. Thus, with this incentive assignment to vertices of $G$, the entire graph is activated. A target vector of cost  ${\sum}_{i=1}^{n}p(u_i)={\sum}_{i=1}^{n}\tau(u_i)-{\sum}_{i=1}^{n}\left[{\sum}_{e\in E(u_i,R)}\omega(e)\right]$ is obtained. Since ${\sum}_{i=1}^{n}\left[{\sum}_{e\in E(u_i,R)}\omega(e)\right]={\sum}_{e\in E(G)}\omega(e)$, the cost of the target vector is ${\sum}_{i=1}^{n}\tau(u_i)-{\sum}_{e\in E(G)}\omega(e)$. From inequality \ref{n3.1}, we conclude that the obtained solution is optimal.
\end{proof}

\noindent In any $(G,\omega)$ denote $\mu(G,\omega)=\min\{\omega(e): e\in E(G) \}$.

\begin{lemma}\label{lemma1}
Let $(G,\omega,\tau)$ be given, where $G$ is connected on $n$ vertices. Assume that for each $u\in V(G)$, $\tau(u)\geq{\sum}_{e\in E(u,G)}\omega(e)-\mu$, where $\mu=\mu(G,\omega)$. Assume that there exists $x\in V(G)$ such that $\tau(x)\geq{\sum}_{e\in E(x,G)}\omega(e)$. Then $\tau$ is degenerate.
\end{lemma}

\noindent \begin{proof}
\noindent We use an induction on $n$. The assertion holds trivially for $n\leq 2$. Let $G$ be a weighted connected graph on $n$ vertices, such that there is a vertex $x\in V(G)$ satisfying $\tau(x)\geq {\sum}_{e\in E(x,G)}\omega(e)$. Let $H$ be a connected components in $G\setminus x$ and $z\in V(H)$ a neighbor of $x$ in $H$. We have
\begin{align}
\tau(z)\geq{\sum}_{e\in E(z,G)}\omega(e)-\mu \geq{\sum}_{e\in E(z,G)}\omega(e)-\omega(xz)={\sum}_{e\in E(z,H)}\omega(e).
\end{align}
Therefore, $H$ satisfies the induction hypothesis and then $\tau$ is degenerate for $H$.
$\tau$ is degenerate for any component of $G\setminus x$ and hence for $G$ itself.
\end{proof}

\noindent Now, by Lemma \ref{lemma1} and Proposition \ref{prop2}, we prove the following theorem.

\begin{thm}
OTVW$\left(\tau(u)\in\left\lbrace{\sum}_{e\in E(u,G)}\omega(e)-\mu(G,\omega),~{\sum}_{e\in E(u,G)}\omega(e)\right\rbrace  \right)$ can be solved in polynomial time.	
\end{thm}

\noindent \begin{proof}
\noindent Let $(G,\omega,\tau)$ be an input of the problem, where $G$ is connected graph. Write $\mu=\mu(G,\omega)$.
If there exists a vertex $x\in V(G)$ such that $\tau(x)\geq{\sum}_{e\in E(x,G)}\omega(e)$,
then by Lemma \ref{lemma1}, $\tau$ is degenerate assignment and the proof is complete by Proposition \ref{prop2}. Otherwise, we have $\tau(u)={\sum}_{e\in E(u,G)}\omega(e)-\mu$, for each $u\in V(G)$. Let $\mathbf{p}^{\ast}$ be an
optimal target vector for $G$ and let $w$ be a last vertex in the activation process corresponding to $\mathbf{p}^{\ast}$.
Vertex $w$ is activated by ${\sum}_{e\in E(u,G)}\omega(e)-\mu$ influence from its neighbors hence there exists an edge
say $e_0$ incident to $w$ which is not used in the activation process.
It follows that sizes of OTV for $G$ and $H=G\setminus e$ (with same weights and thresholds) are equal. Also we have
${\sum}_{e\in E(u,G)}\omega(e)-\omega(e_0)\geq \tau(w)$. Then $\omega(e_0)=\mu$.
Let $e_0=uw$. Let $H_w$ (resp.
$H_w$) be the connected component of $H$ containing $w$ (resp. $u$). Note that possibly $H_u=H_w$. We have
$\tau(w)\geq {\sum}_{e\in E(w,H_w)} \omega(e)$ and $\tau(u)\geq {\sum}_{e\in E(u,H_u)} \omega(e)$. Then the connected graphs
$H_u$ and $H_w$ satisfy the conditions of Lemma \ref{lemma1}, hence by this lemma $\tau$ is degenerate for $H_u$ and $H_w$.
Now by Proposition \ref{prop2}, $OTVW(H_u)$ and $OTVW(H_w)$ and hence $OTVW(H)$ are determined in polynomial time.
\end{proof}

\noindent In the following we show that OTVW has polynomial time solution on graphs $G$, where for each $u\in V(G)$, $\tau(u)={\sum}_{e\in E(u,G)}\omega(e)$ or $\tau(u)=\mu(G,\omega)$.

\begin{thm}
OTVW$\left(\tau(u)\in\left\lbrace{\sum}_{e\in E(u,G)}\omega(e),~\mu(G,\omega)\right\rbrace \right)$ has polynomial time solution.	
\end{thm}

\noindent \begin{proof}
\noindent Assume that $\tau$ is such that for each $u\in V(G)$, $\tau(u)=\mu$ or $\tau(u)={\sum}_{e\in E(u,G)}\omega(e)$. We partition the vertices of $G$ into two sets $V_1$ and $V_2$ as follows. $V_1$ contains vertices whose threshold is $\mu$ and $V_2=V(G)\setminus V_1$. Now, from $G$, we construct a weighted multigraph $\mathbb{M}$
as follows. For each connected component $C$ of $G[V_1]$, we replace $V(C)$ by a single vertex $c$ (representing the component $C$) with threshold $\mu$. Let $C'$ be a set consisting of all such vertices $c$. Define $V(\mathbb{M})=C'\cup V_2$. We put no edge between any two vertices in $C'$, i.e. $C'$ forms an independent set in
$\mathbb{M}$. For each vertex $x\in V_2$ in $G$ and each vertex
$z$ in a connected component $C$ of $G[V_1]$, if there is an edge between $x$ and $z$ in $G$ with weight $k$, we place an
edge of weight $k$ between $x$ and $c$ in multigraph $\mathbb{M}$. Note that if $x$ has $f$ neighbors in $C$ then the edge $xc$ has multiplicity $f$ in $\mathbb{M}$. We do the above process for each vertex of $V_2$ and each connected component of $G[V_1]$ to complete the construction of $\mathbb{M}$.

\noindent Next, we convert $\mathbb{M}$ into a simple graph $F$ as follows. Instead of every edge $xy$ of $\mathbb{M}$, we place a new vertex $s$ with $\tau(s)=\omega(xy)$ and edges $xs$ and $sy$ with $\omega(xs)=\omega(sy)=\omega(xy)$.
Denote by $S$ the set of vertices added in this step. We have the partition $V(F)=C' \cup V_2 \cup S$. Recall that $V_2$ consists of all vertices $u$ such that $\tau(u)={\sum}_{e\in E(u,F)}\omega(e)$. Observe that for each $u\in V_2$,
$\tau(u)={\sum}_{e\in E(u,F)}\omega(e)={\sum}_{e\in E(u,G)}\omega(e)$. Every vertex $s$ in $S$ has degree two and then ${\sum}_{e\in E(s,F)}\omega(e)=2\tau(s)$.
%Also, each vertex of $S$ has exactly one neighbor in $V_2$.
Note that $C'$ contains any vertex of threshold $\mu$ in $F$.

\noindent Now, we organize an ordering $R$ of the vertices in $F$ in such a way that at the beginning of the order, the vertices of $C'$, then the vertices of $S$, and finally the vertices of $V_2$ appear.
Then $R: C', S, V_2$ is the order of vertices in $F$ from left to right. We show that $R$ is a degeneracy ordering in $F$.
We know that $C'$ is an independent set in $F$ and for each $v\in C'$, $\tau(v)=\mu$ and then
$\tau(v)>{\sum}_{e\in E(v,R)}\omega(e)=0$. Now we check the vertices of $S$.
For each $s\in S$, only one edge say $e_0$ is between $C'$ and $s$. We have $\tau(s)=\omega(e_0)={\sum}_{e\in E(s,R)}\omega(e)$.
Finally, for each $v\in V_2$ we have $\tau(v)={\sum}_{e\in E(v,F)}\omega(e)\geq{\sum}_{e\in E(v,R)}\omega(e)$. Therefore, $R$ is a degeneracy order of the vertices in $F$. We converted the input instance $G$ of $OTVW\left(\tau(u)\in\left\lbrace{\sum}_{e\in E(u,G)}\omega(e), \mu \right\rbrace  \right) $ into an instance $F$ of
$OTVW(degenerate)$. According to Proposition \ref{prop2}, there is an activation process on order $R$ by which
the optimal target vector for graph $F$ is obtained.

\noindent Now, we show that the activation process according to $R$ also results in an optimal target vector for $G$.
Every vertex in $C'$ represents a connected component in $G[V_1]$, where all vertices have threshold $\mu$. Thus, in
this activation process, by activating only one vertex from each connected component, the whole component  is activated. By the above definition, instead of every edge $(x,y)\in E(\mathbb{M})$, where $x\in C'$ and $y\in V_2$, the vertex $s$ with $\tau(s)=\omega(x,y)$ is placed, and two edges $(x,s)$ and $(s,y)$ are created with  $\omega(x,s)=\omega(s,y)=\omega(x,y)$. Also, for each $s\in S$ we have $\tau(s)={\sum}_{e\in E(s,R)}\omega(e)$.
Thus, by activating the vertices of $C'$, each vertex $s\in S$ that is placed instead of an edge $(x,y)$, is
activated without receiving incentives. Also, vertex $s$ influences $y$ as much as the weight of edge $(x,y).$ In other words, in order to activate the vertices of $V_2$ in this activation process, we don't need the vertices of $S$.
Now, due to the structure of $F$, for  each vertex $v\in V_2$, $\tau(v)={\sum}_{e\in E(v,F)}\omega(e)={\sum}_{e\in E(v,G)}\omega(e)$. Thus, the activation process on order $R$ assigns the same incentives to the vertices of $V_2$ for both $F$ and $G$. We conclude that, the activation process in which an optimal solution is obtained for $F$ also gives the optimal solution for $G$.
\end{proof}

\section{Concluding remarks}

\noindent A complete model for spread of influence in networks is to consider bi-directed graphs with weighted edges. Let $(\overleftrightarrow{G}, \omega)$ be a weighted bi-directed graph in which for any two vertices $u$ and $v$ in $\overleftrightarrow{G}$, the influence of $u$ on $v$ (resp. the influence of $v$ on $u$) is proportional to $\omega(uv)$ (resp. $\omega(vu)$). In real world networks too, influence of two people on each other are not necessarily identical. For example very influential persons have much influence on their neighbors than ordinary people. In these situations we need to use bi-directed edge-weighted graphs. Consider a triple
$(\overleftrightarrow{G},\omega,\tau)$, where $\tau$ assigns a threshold $\tau(u)\leq d^{in}(u)$ to each vertex $u$. The activation process corresponding to $(\overleftrightarrow{G},\omega,\tau)$, is defined as follows. Initially a subset $A_0$ of vertices in $\overleftrightarrow{G}$ is activated. $A_t$ is the set of vertices activated in round $t$. Then a vertex $x\in V(\overleftrightarrow{G})\setminus A_t$ becomes active in round $t+1$ if and only if
${\sum}_{e\in E_{t}^{in}(x)}\omega(e)\geq\tau(x)$, where $E_t^{in}(x)$ is the set of incoming edges from $A_t$ to $x$.

\noindent Target sets, target vectors and optimal target sets in $(\overleftrightarrow{G},\omega,\tau)$ and any directed graph with weighted edges
are defined similarly. It was proved in \cite{AZ} that optimal target vectors in bi-directed edge-weighted paths and cycles can be solved polynomially by dynamic programming algorithms. By TSSWD we mean the problem of determining optimal size of target sets
in directed or bi-directed graphs. Note that every instance $(G=K_n, \omega, \tau)$ of TSSW(complete) can easily be
transformed to a bi-directed complete graph $H$. It is enough to replace each edge $e=uv$ of weight $\omega(uv)$ in $G$ by
two directed edges $(u,v)$ and $(v,u)$ with new weights $\omega'(u,v)=\omega'(v,u)=\omega(uv)$ to obtain $H$. It follows
that TSSWD is $\NP$-complete for bi-directed complete graphs. This problem is non-trivial
if we only consider unilateral directed graphs, where between any two vertices $u$ and $v$, either there is no edge from
$u$ to $v$ or there is no edge from $v$ to $u$. It was proved in \cite{KSZ2} that to determine optimal target sets in bilateral directed graphs with constat thresholds $2$ is $\NP$-complete. Clearly, a directed counterpart of TSSW(complete)
problem is to consider tournaments. A tournament on $n$ vertices is any edge orientation of simple complete graph $K_n$.
Define similarly TSSWD(tournament) as directed counterpart of TSSW(complete). By combining the proof of Proposition
\ref{prop1} and $\NP$-completes of optimal target sets in bilateral directed graphs in \cite{KSZ2} we obtain the following
which we present without proof details.

\begin{prop}
TSSWD(tournament) i.e. TSSWD for tournaments is \NP-complete, where the weight of every directed edge is positive.
\end{prop}

\end{document}